\documentclass[twocolumn,showpacs,aps,prd,amsmath,amssymb,amsfonts%
]{revtex4-1}
\usepackage{graphicx}
\usepackage{bm}
\usepackage{relsize}

\begin{document}
\newcommand*{\ea}{\textit{et al.}}
\newcommand*{\zpc}[3]{Z.~Phys.~C \textbf{#1}, #2 (#3)}
\newcommand*{\plb}[3]{Phys.~Lett.~B \textbf{#1}, #2 (#3)}
\newcommand*{\phrd}[3]{Phys.~Rev.~D~\textbf{#1}, #2 (#3)}
\newcommand*{\phrl}[3]{Phys.~Rev.~Lett.~\textbf{#1}, #2 (#3)}
\newcommand*{\pr}[3]{Phys.~Rev.~\textbf{#1}, #2 (#3)}      
\newcommand*{\npbps}[3]{Nucl.~Phys.~B (Proc. Suppl.) \textbf{#1}, #2 (#3)}  
\newcommand*{\rmph}[3]{Rev.~Mod.~Phys.~\textbf{#1}, #2 (#3)}
\newcommand*{\ibid}[3]{\textit{ ibid.} \textbf{#1}, #2 (#3)}
\newcommand*{\epjc}[3]{Eur. Phys. J. C \textbf{#1}, #2 (#3)}
\newcommand*{\nl}{\nonumber \\}
\newcommand*{\bea}{\begin{eqnarray}}
\newcommand*{\eea}{\end{eqnarray}}
\newcommand*{\bi}{\bibitem}
\newcommand*{\be}{\begin{equation}}
\newcommand*{\ee}{\end{equation}}
\newcommand*{\eg}{e.g.}
\newcommand*{\ra}{\rightarrow}
\newcommand*{\rf}[1]{(\ref{#1})}
\newcommand*{\dek}[1]{\times10^{#1}}
\newcommand*{\lag}{{\mathcal L}}
\newcommand*{\tff}[2]{{F(#1,#2)}}
\newcommand*{\qso}{{Q_1^2}}
\newcommand*{\qst}{{Q_2^2}}
\newcommand*{\fai}{\bm\phi}
\newcommand*{\vrp}{{V\!\rho\pi}}
\newcommand*{\orp}{{\omega\rho\pi}}
\newcommand*{\prp}{{\phi\rho\pi}}
\newcommand*{\op}{{\omega^\prime}}
\newcommand*{\opp}{{\omega^{\prime\prime}}}
\newcommand*{\die}{{{e^+e^-}}}
\def\babar{\mbox{\slshape B\kern-0.1em{\smaller A}\kern-0.1em
    B\kern-0.1em{\smaller A\kern-0.2em R}}}

\title{Vector meson dominance and the $\bm{\pi^0}$ transition form factor}

\author{Peter Lichard}
\affiliation{
Institute of Physics, Silesian University in Opava, 746 01 Opava, Czech Republic\\
and\\
Institute of Experimental and Applied Physics, Czech Technical University in Prague, 128 00 Prague, Czech Republic
}

\begin{abstract}
It is shown that the $\pi^0$ transition form factor $F(Q_1^2,Q_2^2)$ differs
substantially from its one-real-photon limit $F(Q_1^2,0)$ even for rather
small values of $Q_2^2$ ($\approx 0.1~\mathrm{GeV}^2$), which cannot be
excluded in experiments with one ``untagged'' electron. It indicates that
the comparison of data with theoretical calculations, which usually 
assume $Q_2^2=0$, may be untrustworthy.  
Our phenomenological model of the $\pi^0$ transition form factor is based on 
the vector-meson-dominance hypothesis and all its
parameters are fixed by using the experimental data on the decays of
vector mesons. The model soundness is checked in the two-real-photon limit, 
where it provides a good parameter-free description of the $\pi^0\ra2\gamma$
decay rate, and in the $\pi^0$ Dalitz decay. The dependence of 
$F(Q_1^2,Q_2^2)$ on $Q_1^2$ at several fixed values of $Q_2^2$ is presented 
and the comparison with existing data performed. 
\end{abstract}

\maketitle
The issue of the $\pi^0$ transition form factor has recently attracted
renewed interest in connection with the precise measurements of the 
\babar~Collaboration \cite{babar}, which seem to indicate that the asymptotic
limit predicted by perturbative QCD \cite{lepage80} has been exceeded. 
A comprehensive review of the current theoretical situation with an extended
list of references can be found in Ref.~\cite{arriola10}. On a 
phenomenological side, it has recently been shown \cite{jurlic} that the 
vector-meson-dominance (VMD) hypothesis \cite{vmd,gellmann1961,gellmann1962} 
leads to a correct description of the two-photon decay of the
$\pi^0$ if the parameters are fixed by the data on the partial decay widths of 
the vector mesons \cite{note1}.
The rate of this decay is related to the real-photon limit \tff{0}{0}~of the
$\pi^0$ transition form factor. It is therefore tempting to use the VMD also for
the construction of the $\pi^0$ transition form factor, which parametrizes the 
dynamics of the process in which two off-mass-shell photons fuse and form 
a $\pi^0$.

The experimental data on the $\pi^0$ transition form factor are taken in the 
process $e^-+e^-(e^+)\ra
e^-+\pi^0+e^-(e^+)$, where the photon virtualities are given by the
electron ($e^+$ or $e^-$) momenta transfer squared, $Q_i^2=-q_i^2$,
$i=1,2$.
To get the one-real-photon transition form factor only the data sample is
utilized in which one lepton exhibits a small momentum transfer, \eg, 
$|q^2|<0.18~\mathrm{GeV}^2$ in the latest \babar~experiment \cite{babar}.

The transition form factor $\tff{\qso}{\qst}$ is defined by its appearance
in the $\gamma^*\gamma^*\pi^0$ vertex, which describes the fusion of two 
virtual photons with the four-momenta $q_1$ and $q_2$ into a $\pi^0$
\be\label{definition}
T_{\mu\nu}=-ie^2\epsilon_{\mu\nu\alpha\beta}q_1^\alpha 
q_2^\beta \tff{\qso}{\qst},
\ee
where $e$ is the elementary electric charge. This definition of the
transition form factor agrees with that used in \cite{babar,cleo1998}, but
differs from that in \cite{cello1991}, where the factor $e^2$ 
was absorbed in $\tff{\qso}{\qst}$. The two-real-photon value of the pion 
transition form factor is related to the two-photon decay width of the 
$\pi^0$ by
\be\label{twogamma}
\Gamma(\pi^0\to 2\gamma)=\frac{\pi\alpha^2}{4}m_\pi^3 F^2(0,0),
\ee
where $\alpha$ is the fine-structure constant. 

Another process in which the
transition form factor plays a role is the decay $\pi^0\to e^+e^-\gamma$, 
which was suggested by Dalitz \cite{dalitz51} as an explanation of the anomalous events recorded in photographic 
emulsions exposed in high-flying balloons \cite{carlson50}. The original 
evaluation of the branching ratio 
\be\label{bdalitz}
B=\frac{\Gamma(\pi^0\to e^+e^-\gamma)}{\Gamma(\pi^0\to 2\gamma)}
\ee
by Dalitz as well as a later one \cite{krollvada55}, did not consider 
a possible form factor. The latter was included in \cite{berman1960}. 
The differential branching ratio \rf{bdalitz} in the Berman--Geffen 
\cite{berman1960} variable $x=M^2/m^2_\pi$, where $M$ is the mass of 
the $e^+e^-$ pair, reads as 
\be\label{dalitzx}
\frac{dB}{dx}=\frac{2\alpha}{3\pi}\frac{(1-x)^3}{x}
\left(1+\frac{2\epsilon}{x}\right)\sqrt{1-\frac{4\epsilon}{x}}\ F_D^2(x),
\ee
where $\epsilon=m^2_e/m^2_\pi$. Form factor $F_D$ is related to the $\pi^0$ 
transition form factor by
\be\label{dalitzff}
F_D(x)=\frac{F(-xm^2_\pi,0)}{F(0,0)}.
\ee
The total branching ratio \rf{bdalitz} is not very sensitive to the shape
of the form factor \rf{dalitzff} and cannot serve as a stringent test of
theoretical calculations or phenomenological models. Even the original 
Dalitz formula, which did not include the form factor at all, leads to 
$B=1.185\%$, which agrees with the experimental value of $(1.188\pm0.035)\%$ 
\cite{pdg2010}. The form factor \rf{dalitzff} is at small $x$ usually 
parametrized as
\[
F_D(x)=1+ax.
\]
In 1961, Gell-Mann and Zachariasen \cite{gellmann1961} showed that the form 
factor $F_D$  is dominated by two resonances, namely $\rho$ and $\omega$,  
and got a positive $a$ equal to $m_\pi^2(m_\rho^{-2}+m_\omega^{-2})/2$, 
in agreement with today's observations. Until late 1980's, most 
experiments had indicated negative values of $a$, see \cite{fonvieille1989}
and a list of preceding experimental results there.

\begin{figure}[h]
\includegraphics[width=8.6cm]{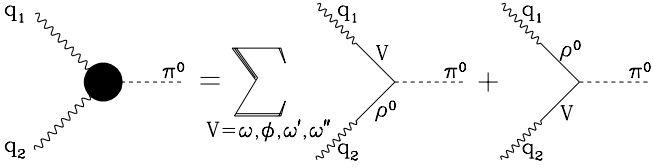}
\caption{\label{fig:fig1}Feynman diagrams defining our model of the
$\pi^0$ transition form factor.}
\end{figure}

Our model of the pion transition form factor $\tff{\qso}{\qst}$ is defined 
in Fig.~\ref{fig:fig1}. In addition to the
$\rho\omega$ intermediate state considered in 
\cite{jurlic,gellmann1961,gellmann1962}, 
we include the following companions 
of the $\rho^0$: $\phi$, $\omega(1420)$ (denoted as $\op$ in what follows), 
and $\omega(1650)$ (denoted as $\opp$).
All these resonances, except $\opp$, have recently been considered in the 
model of the one-real-photon transition form factors of $\pi^0$, $\eta$,
and $\eta^\prime$ \cite{dubnickova2004}.

The combination of an isoscalar resonance with an isovector one in 
the intermediate states is unique for the $\pi^0$ transition form factor
and implies the equal contribution of the $I=0$ and $I=1$ currents to
$F(0,0)$. With the $\eta$ and $\eta^\prime$, the situation is different.

In order to evaluate the Feynman diagrams depicted in Fig.~\ref{fig:fig1},
we use the Lagrangian
\be
\label{lagvrhpi}
{\cal L}_{\vrp} =
G_\vrp\epsilon_{\mu\nu\alpha\beta}
\left(\partial^\mu V^\nu\right)\left(\fai\cdot
\partial^\alpha\bm{\rho}^\beta\right),
\ee
where $V$ is the operator of a neutral vector meson field, $\bm\rho$ is 
the isovector of the $\rho$ meson fields, and $\fai$ is that of the pion 
fields. The coupling of photons to neutral vector mesons is given by the 
VMD Lagrangian (a little nonstandard notation of \cite{tripi} adopted from
\cite{donnell} is used)
\be  \label{lage}
{\cal L}_{\rm em} =
-\sum_V\frac{eg_{V\gamma}}{2}m_V^2A^\mu V_\mu,
\ee
where $A^\mu$ is the electromagnetic field operator.

After comparing the amplitude corresponding to Feynman diagrams in
Fig.~\ref{fig:fig1} with the definition of the transition 
form factor \rf{definition} we extract the latter in the following
form 
\bea\label{tff}
\tff{\qso}{\qst}&=&\frac{g_{\rho\gamma}}{4}\sum_{V=\omega,\phi,\ldots}
G_V\left[R_V(\qso)R_\rho(\qst)\right.\nl
&+&\left. R_V(\qst)R_\rho(\qso)\right],
\eea
where $G_V=G_{\vrp}g_{V\gamma}$ and functions
\be\label{rv}
R_V(Q^2)=\frac{m_V^2}{m_V^2+Q^2}
\ee
are the scalar parts of the vector meson propagators below the physical cut
threshold in $s=-Q^2$, $s_\mathrm{th}=m_\pi^2$. Tensor parts do not
contribute thanks to the presence of the Levi-Civita tensor in
\rf{lagvrhpi}. For $R_\rho(Q^2)$ we will alternatively use the form
\be\label{rrhorun}
R_\rho(Q^2)=\frac{M_\rho^2(0)}{M_\rho^2(-Q^2)+Q^2}
\ee
with the running mass squared given by the dispersion formula
\be\label{running}
M_\rho^2(s)=M_\rho^2(0)-\frac{s}{\pi}{\mathcal P}\!\!\!\!\int
\limits_{m_\pi^2}^\infty
\frac{m_\rho\Gamma_\rho(s^\prime)}{{s^\prime}(s^\prime-s)}ds^\prime
\ee
and satisfying $M_\rho^2(m_\rho^2)=m_\rho^2$ and $dM_\rho^2/ds=0$ at
$s=m_\rho^2$. The energy dependent total
width $\Gamma_\rho$ includes the contributions from the following final
states: $\pi^+\pi^-$, $K\bar{K}$, $\pi^0\omega$, $\eta\pi\pi$,
$\pi^0\gamma$, $\eta\gamma$, and $\pi^+\pi^-\gamma$. For details, 
see \cite{running}.

Inserting \rf{tff} into \rf{dalitzff} and calculating the derivative of 
$F_D(x)$ at $x=0$ we are getting the following expression for the slope
of the Dalitz decay form factor
\be\label{mya}
a=\frac{m_\pi^2}{2}\left[\frac{1}{M_\rho^2(0)}\left(1-D\right)+
\frac{1}{\sum_V G_V}\sum_V\frac{G_V}{m_V^2}\right].
\ee
Here, $D$ is the derivative of the running mass \rf{running} at $s=0$.
When a fixed mass form \rf{rv} is used also for $R_\rho(Q^2)$, then $D=0$
and $M_\rho^2(0)=m_\rho^2$. If we, in addition, keep only the first term
$V=\omega$ in the sums over $V$ above, we recover the Gell-Mann--Zachariasen 
\cite{gellmann1961} formula.

The values of coupling constants $G_{\vrp}$ are determined as follows.
For the vector mesons with masses below or close to the $\rho\pi$ threshold 
[$\omega(782)$, $\phi(1020)$] we consider the decay chain 
$V\ra\rho+\pi\ra 3\pi$. The comparison of the decay width formula given in 
\cite{tripi} with the recommended value from \cite{pdg2010} yields 
the product $G_{\vrp}^2g_\rho^2$, where $g_\rho$ is the coupling constant
of the usual $\rho\pi\pi$ interaction Lagrangian. Using the value 
$g_\rho^2=35.70\pm0.19$, as it follows from the $\rho$ meson decay width
\cite{pdg2010}, we get $G_{\vrp}^2$.
For the vector mesons with higher masses [$\op$, $\opp$, $J/\psi(1S)$] it
is sufficient to explore the simpler $V\ra\rho+\pi$ decay width formula. 
Concerning the $\op$, the Review of Particle Physics \cite{pdg2010} 
gives only intervals for its mass and total width and no quantitative 
estimate for the $\rho\pi$ branching fraction. We use therefore the 
values \cite{note2}
$m_{\op}=(1.38\pm0.02\pm0.07)$~GeV/$c^2$ and 
$\Gamma_{\op}=(0.13\pm0.05\pm0.10)$~GeV as measured by the {\babar} 
Collaboration \cite{babar2007} and $B(\op\ra\rho+\pi)=(69.9\pm2.9)\%$ 
from the wavelet analysis \cite{henner02} of the $\die$ annihilation data. 
In the case of the $\opp$, we again use the {\babar} \cite{babar2007} values
$m_{\opp}=(1.667\pm0.013\pm0.006)$~GeV/$c^2$ and 
$\Gamma_{\opp}=(0.222\pm0.025\pm0.020)$~GeV 
with the wavelet analysis \cite{henner02} branching fraction 
$B(\opp\ra\rho+\pi)=(38.0\pm1.4)\%$. The resulting
values of the coupling constants $G_{\vrp}$ squared are summarized 
in Table \ref{tab:couplings}. The coupling constants $G_{\orp}$ and $G_{\prp}$ 
themselves differ in sign, as it follows from the analysis of the 
$\rho\ra\pi\gamma$ decay \cite{tripi} and from the SU(3) symmetry 
\cite{donnell}. This results in the negative sign of $G_\phi$ shown in
Table \ref{tab:couplings}.
\begin{table}
\caption{\label{tab:couplings}The squares of the coupling constants in
Lagrangians \rf{lagvrhpi} and \rf{lage} obtained from the vector meson 
decay data described in the text. Also shown are the parameters $G_V$, 
which enter the transition form factor \rf{tff}.}
\begin{ruledtabular}
\begin{tabular}{cccc}
$V$  & $G^2_{\vrp}~(\mathrm{GeV}^{-2})$ & $g^2_{V\gamma}\dek{2}$ &
$G_V~(\mathrm{GeV}^{-1})$\\
\colrule
$\omega(782)$  &$216.2\pm3.0$    &$1.375\pm0.046$  & $~~1.724\pm 0.031$\\
$\phi(1020)$   &$0.676\pm0.020$  &$2.214\pm0.031$  & $-0.122\pm 0.002$\\
$\omega(1420)$ &$11.7\pm1.1 $    &$0.20\pm0.17$  & $~~0.152\pm0.136$\\
$\omega(1650)$ &$3.97\pm0.61$    &$0.76\pm 0.11$  & $~~0.174\pm0.026$
\end{tabular}
\end{ruledtabular}
\end{table}

Now, we determine the coupling constants in the VMD Lagrangian \rf{lage}.
The value $g^2_{\rho\!\gamma}=4/g_\rho^2=0.1120(10)$ follows from the 
normalization of the charged pion form factor. The squares of other coupling 
constants $g_{V\!\gamma}$ are evaluated from the dilepton decay width of the 
corresponding vector mesons. In the case of the $\omega$, the $\die$ decay 
width from \cite{pdg2010} is used. For the $\phi$, we obtain it as a product 
of the full width and the $\die$ branching fraction. To get 
$g^2_{\opp\!\gamma}$, we combine the full width from \cite{babar2007} (shown
above) with the $\die$ branching fraction $B(\opp\ra\die)=(32\pm1)\dek{-7}$
from \cite{henner02}. Let us note that the product of the $\rho\pi$ and 
$\die$ branching fractions from 
\cite{henner02} 
agrees with $B(\opp\ra\die)B(\opp\ra\rho+\pi)=(1.3\pm0.1\pm0.1)\dek{-6}$
got later by the \babar~Collaboration \cite{babar2004}. The same concerns 
the full width. Unfortunately, the situation with the $\op$ is more 
controversial. The product of the $\rho\pi$ and $\die$ branching fractions
from the wavelet analysis \cite{henner02} is about twofold of that
observed by \babar~\cite{babar2004}. In order not to overestimate the
contribution of the $\op$ to the $\pi^0$ transition form factor, we use
the \babar~value $B(\op\ra\die)B(\op\ra\rho+\pi)=(0.82\pm0.05\pm0.06)\dek{-6}$
to get $G^2_{\op}=G_{\op\!\rho\pi}^2g_{\op\!\gamma}^2$, from which the estimate
of $g_{\op\!\gamma}^2$ presented in Table~\ref{tab:couplings} is obtained. 

\begin{figure}[h]
\includegraphics[width=8.6cm]{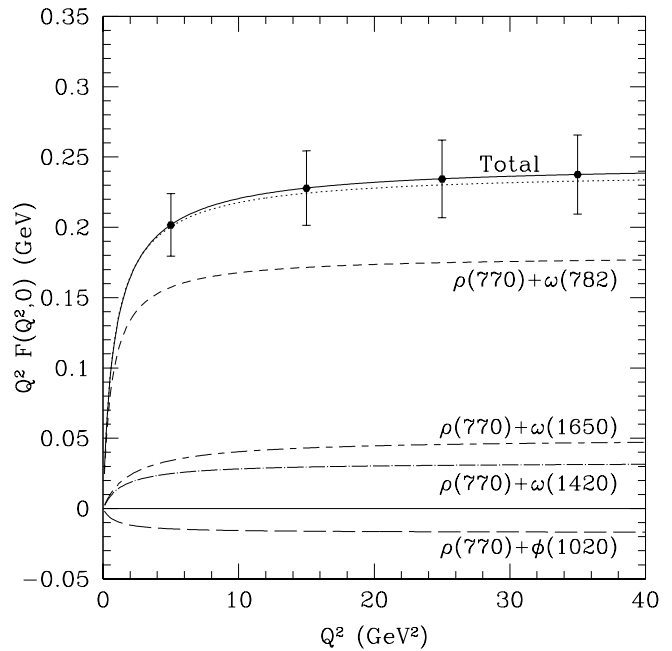}
\caption{\label{fig:fig2}The individual contributions to the pion transition 
form factor $F(Q^2,0)$ multiplied by $Q^2$ and their sum (solid curve)
with the uncertainties originating in the errors of input parameters.
The sum of individual contributions (not shown) that were calculated assuming 
the constant mass of the $\rho(770)$ is depicted by the dotted curve.}
\end{figure}

\begin{figure}
\includegraphics[width=8.6cm]{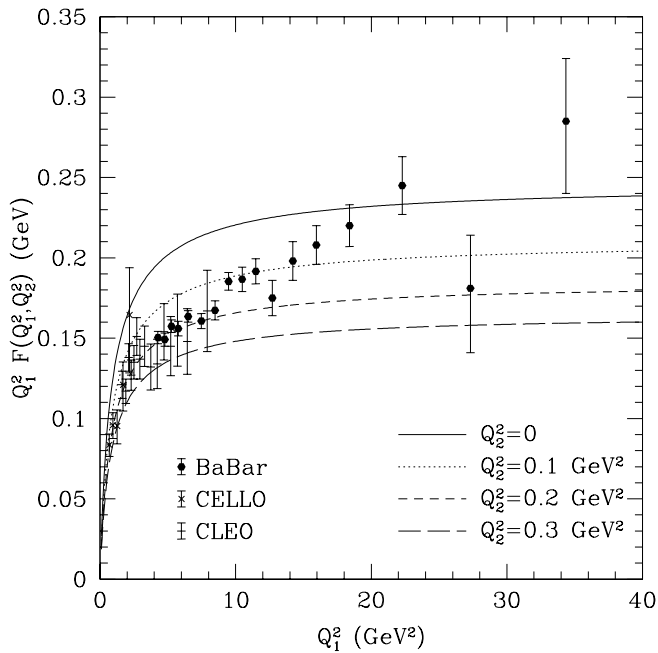}
\caption{\label{fig:fig3}Dependence of $Q^2_1F(Q_1^2,Q_2^2)$ on $Q_1^2$
for four fixed values of $Q_2^2$ calculated in the full version of our
model. The CELLO \cite{cello1991}, CLEO \cite{cleo1998}, and 
\babar~\cite{babar} data are also shown.}
\end{figure}

As all the parameters of our model of the $\pi^0$ transition form factor
are now determined, we are ready to test its soundness by evaluating the 
characteristics of the two-photon and Dalitz decays of the neutral pion 
and comparing them to the measured values. 
To avoid
the cumulation of the errors, we do not use Table \ref{tab:couplings}
for the error analysis, but calculate the errors of the results directly 
from the input data (masses, full widths, branching fractions, and their 
errors). To account for possible correlations among the input quantities,
we sum the contributions to the final errors from various sources
linearly. Two of the calculated quantities, namely the mean lifetime
of $\pi^0$ and the Dalitz decay slope parameter are shown in
Table~\ref{tab:decays} for various versions of our model, i.e., for
various choices of the isoscalar vector mesons entering the sum in
Eq.~\rf{tff} and for two possible treatments of the $\rho$ propagator. The 
branching ratio of the Dalitz decay to the two-photon decay is not shown 
in the table. It acquires the same value of 1.196(1)\% in all versions 
of the model, in agreement with the experimental value of 1.188(35)\% 
\cite{pdg2010}. 
\begin{table}[h]
\caption{\label{tab:decays}Comparison of various versions of our model with
the $\pi^0$ decay data. When converting the calculated
$\pi^0\ra2\gamma$ decay rate into the $\pi^0$ mean lifetime $\tau$, the experimental
branching fraction of 0.988 was used. The results above the horizontal divider
were obtained with the constant $\rho$ mass, those below it with the
running mass \rf{running}.}
\begin{ruledtabular}
\begin{tabular}{ccc}
$V$  & $\tau\dek{17}$(s) & $a\dek{2}$ \\
\colrule
$\omega$     &$7.6\pm0.4 $&$ 3.002\pm0.002 $\\
$\omega,\phi$&$8.8\pm0.6 $&$ 3.049\pm0.004 $\\
\colrule
$\omega,\phi$  &$8.8\pm0.6$&$ 3.282\pm0.003$\\
$\omega,\phi,\omega^\prime$  &$ 7.3\pm1.9 $&$3.190\pm0.097 $\\
$\omega,\phi,\omega^{\prime\prime} $&$ 7.2\pm0.7 $&$ 3.163\pm0.024 $\\
$\omega,\phi,\omega^\prime,\omega^{\prime\prime} $&$ 6.1\pm1.7 $&
$ 3.089\pm0.099 $\\
\colrule
Data \cite{pdg2010} &$ 8.4\pm0.5 $&$ 3.2\pm0.4$ 
\end{tabular}
\end{ruledtabular}
\end{table}
Our results concerning the $\pi^0$ transition form factor are presented
in Figs.~\ref{fig:fig2} and \ref{fig:fig3}. The contributions
of the various $\pi^0\rho^0V$ vertexes to the one-real-,
one-virtual-photon form factor together with their sum are shown
in Fig.~\ref{fig:fig2}. They were all calculated using the running
mass \rf{running} of the $\rho$ meson. In addition, the sum
of the same contributions but calculated using the fixed $\rho$ mass
is depicted by a dotted curve. It differs from the running-mass case
only marginally.
The strong dependence of the transition form factor $F(Q^2_1,Q^2_2)$ 
on the virtuality $Q^2_2$, which is given by the four-momentum transfer 
of the ``untagged'' electron, is demonstrated in Fig. \ref{fig:fig3}. 
The expression $Q_1^2F(Q_1^2,Q_2^2)$ is presented as a function of $Q_1^2$ 
for four different virtualities $Q_2^2$ from 0 to 0.3~GeV$^2$. The full 
version ($V=\omega,\phi,\op$, and $\opp$) of our model with running mass
of the $\rho$ is used. Comparison with existing data indicates that 
our model would be able to describe them if the (unmeasured, but certainly
nonvanishing) 
absolute value of the momentum transfer squared of the ``untagged'' 
electron decreased with the rising $Q_1^2$. 
A more quantitative account
is given in Table~\ref{tab:fitting}, which shows that the agreement of
our model with the data below $Q_1^2=9$~GeV$^2$ is excellent but
deteriorating for higher $Q_1^2$. 
There are two possible ways of improving our model in an attempt to get a
better description of the high-$Q_1^2$ data. First, the inclusion of
higher isovector [$\rho(1450)$, $\rho(1700)$] and isoscalar [$\phi(1680)$,
$\phi(2170)$] resonances. Second, a consequent use of the running-mass 
propagators also for other resonances [not only for $\rho(770)$ as here], 
which guarantee the correct analytic properties of the transition form 
factor. 

\begin{table}[h]
\caption{\label{tab:fitting}
The comparison of our parameter-free model with the data in various 
$Q_1^2$ ranges. The virtuality $Q_2^2$ of the photon radiated from the 
``untagged'' electron was assumed the same for all $n$ data points within 
a particular $Q_1^2$ range and determined by minimizing the $\chi^2$.
The confidence levels (C.L.) are also shown.}
\begin{ruledtabular}
\begin{tabular}{ccccc}
$Q^2_1$~(GeV$^2$)  
& n & $\chi^2$ & $Q^2_2$~(GeV$^2$)& C.L. (\%) \\
\colrule
~0--9  & 27 & 10.67 & $0.187\pm0.008$ & 99.7 \\
~9--18 & ~6 & ~3.77 & $0.107\pm0.013$ & 58.3\\
18--36 & ~4 & ~4.47 & $0.010\pm0.025$ & 21.5 
\end{tabular}
\end{ruledtabular}
\end{table}

The effect of nonvanishing $Q_2^2$ has already been quantitatively studied 
in terms of
the photon momentum asymmetry parameter $A=(Q_1^2-Q_2^2)/(Q_1^2+Q_2^2)$ 
within the spectral quark model \cite{broniowski09} and the 
incomplete vector-meson dominance (IVMD) \cite{arriola10}. In the former,
the transition form factor decreases with rising $Q_2^2$ (falling $A$)--see 
Fig.~2 in \cite{broniowski09}--as in our model. In the latter, the 
tendency is inverse--see Fig. 7 in \cite{arriola10}. This difference 
is caused by not keeping the IVMD parameters $c$ and $M_V$ constant, 
but allowing them to vary in order to get the best fit for each particular 
$A$.

Our model, in spite of its deficiencies, supports the conclusion of Refs.
\cite{broniowski09,arriola10} that it is important to pay
more attention to the dependence of the transition form factor on both
virtualities in theoretical calculations and experimental analyses.

I thank S. Dubni\v{c}ka, J. Jur\'{a}\v{n}, and J.~Pi\v{s}\'{u}t 
for discussions and W. Broniowski for correspondence. This work was 
supported by the Czech Ministry of Education,
Youth and Sports under Contracts No. LC07050 and No. MSM6840770029.

\end{document}